# Network Functions Virtualization Architecture for Gateways for Virtualized Wireless Sensor and Actuator Networks[+]

Carla Mouradian[¥1], Tonmoy Saha[¥2], Jagruti Sahoo[¥3], Mohammad Abu-Lebdeh[¥4],
Roch Glitho[¥5], Monique Morrow[€1] , Paul Polakos[£1]
[¥]Concordia University, Montreal, Canada,
[€]CISCO Systems, Zurich, Switzerland,
[£]CISCO Systems, New Jersey, USA
[¥1]ca_moura@encs.concordia.ca, [¥2]to_saha@encs.concordia.ca, [¥3]jagrutiss@gmail.com,
[¥4]m_abuleb@encs.concordia.ca, [¥5]glitho@ece.concordia.ca, [€1]mmorrow@cisco.com, [£1]ppolakos@cisco.com

*Abstract*— **Virtualization enables multiple applications to share the same wireless sensor and actuator network (WSAN). However, in heterogeneous environments, virtualized wireless sensor and actuator networks (VWSAN) raise new challenges, such as the need for on-the-fly, dynamic, elastic, and scalable provisioning of gateways. Network Functions Virtualization (NFV) is a paradigm emerging to help tackle these new challenges. It leverages standard virtualization technology to consolidate special-purpose network elements on commodity hardware. This article presents NFV architecture for VWSAN gateways, in which software instances of gateway modules are hosted in NFV infrastructure operated and managed by a VWSAN gateway provider. We consider several VWSAN providers, each with its own brand or combination of brands of sensors and actuators/robots. These sensors and actuators can be accessed by a variety of applications, each may have different interface and QoS (i.e., latency, throughput, etc.) requirements. The NFV infrastructure allows dynamic, elastic, and scalable deployment of gateway modules in this heterogeneous VWSAN environment. Furthermore, the proposed architecture is flexible enough to easily allow new sensors and actuators integration and new application domains accommodation. We present a prototype that is built using the OpenStack platform. Besides, the performance results are discussed.** [2]

*Keywords— Gateway; Network Functions Virtualization; Virtualization; Wireless Sensor and Actuator Networks*

## I. Introduction

Research on sensor network virtualization [1] has become prominent in recent years. Virtualization technology abstracts sensor resources as logical units and allows for their efficient and simultaneous use by multiple applications, even if they have conflicting requirements and goals. New applications can be deployed in the same WSN with minimal efforts. More importantly, reusing the same sensors' capability by multiple applications transforms WSN into a multi-purpose sensing platform in which several virtual WSNs (VWSNs) are created on-demand, each tailored for a specific task or objective. Actuators are often incorporated in WSNs to make more powerful applications, thus the concept of virtualized wireless sensor and actuator network (VWSAN).

Gateways are required for the interactions between applications and heterogeneous, multivendor VWSANs. They are generally complex. Furthermore, it is difficult and expensive to upgrade them when new-brand sensors and actuators/robots are deployed. In addition, their capabilities do not scale when the number of applications and the corresponding workload in VWSANs change dynamically.

Network Functions Virtualization (NFV) [2] is an emerging paradigm in overcoming the aforementioned challenges. NFV permits standard virtualization technology to consolidate dedicated network elements (e.g., firewalls, network address translation (NAT)) onto commodity hardware. By implementing network functions as software instances called virtual network functions (VNFs), NFV reduces the operational costs and provides hardware independence. Moreover, on-the-fly, dynamic, scalable, and elastic provisioning of network services are among its benefits.

This paper presents an NFV architecture for Virtualized Wireless Sensor and Actuator Networks Gateways (VWSAN). The firmware/hardware used to provide VWSAN Gateway functionalities are replaced by VNFs deployed in an NFV infrastructure. We enable a granular provisioning of NFV, such as decomposing the gateway into fine-grained modules – e.g., protocol converter, information model converter, etc. – to be implemented as VNFs. More importantly, granular NFV is best suited for virtualized WSANs, wherein the dynamic growth in the number of applications and addition of new-brand sensors require a rapid introduction of new VNFs and update of existing VNFs. VNFs are instantiated on-the-fly and chained to realize a service in VWSAN.

The architecture introduces a new business actor - the VWSAN Gateway Provider – in addition to the traditional

---

actors, meaning the Application Provider and the VWSAN Provider. This new actor plays a dual role. On the one hand, it provides the VNFs, chained to make on-the-fly gateways. On the other hand, it operates and manages the infrastructure in which the VNFs are executed. We acknowledge that the introduction of this new actor does bring a host of additional security and trustability challenges. We consider these challenges outside the scope of this paper. More and more standardization work will certainly be required to enable secure and trustable interactions between different NFV actors, as the business model opens up.

The next section introduces a motivating scenario, requirements, and discusses state-of-the-art. The proposed architecture is presented in Section 3, followed by the implementation details, the prototype, and performance results in Section 4. In the last section, we conclude the paper and outline future work.

## II. CRITICAL OVERVIEW OF STATE-OF-THE-ART

### A. Motivating Scenario

The ability of sensors to withstand harsh environments makes WSAN a potential tool for forest monitoring and protection. The use of WSAN allows forest researchers to understand the impact of air pollutants (e.g., $CO_2$, ozone, etc.) and climate change on tree growth. We consider a potential scenario in which a forest monitoring agency collects environment data using sensor infrastructure provided by a third party VWSAN provider. The sensors are of various capabilities, including temperature, humidity, rain gauge, CO2 detector, and wind speed sensors. Let us also consider a wildfire management agency that needs to be promptly notified when a fire occurs in the forest so that it can deploy a fleet of heterogeneous fire fighter robots to suppress the wildfire. WSAN virtualization would allow for the concurrent execution of the forest monitoring and wildfire management applications on the same sensors. In order to collect measurements from the sensors and send commands to the robots in a heterogeneous environment, a gateway is needed for the interactions between the application domain and WSAN domain.

### B. Requirements

First, the gateway must support *standard northbound and proprietary southbound interfaces*. An example of standard interface could be the widely used Sensor Markup Language (SenML) [3] carried over HTTP. It is designed to encode sensor measurements and device parameters. The proposed architecture must be *extensible* to support future scenarios and new application domains. In addition, the architecture must be *elastic* to allow for the efficient utilization of underlying physical resources. The architecture must be *scalable* to promote the accelerated growth of the number of applications.

The architecture should also provide at least two key gateway functions: *Protocol conversion* and *information model conversion*.

The architecture must ensure that the execution of gateway modules achieves *performance* similar to when they are executed in a traditional WSAN gateway. In particular, the performance metrics that require significant attention are latency, throughput, and overhead. The NFV architecture must be *flexible* enough to support the integration of various brands sensors, and it must have the *ability to support different business models*.

### C. The State-of-the-Art and Its Shortcomings

Our motivating scenario closely resembles the WSN and Internet of Things (IoT) scenarios, which involve a broad range of sensors, IoT devices, and communication technologies at the IoT device domain (e.g., 6LoWPAN, ZigBee, Bluetooth, etc.) and network domain (e.g., 2G/3G, LTE, LAN, etc.). In state-of-the-art for WSN/IoT gateway architectures, the main focus has been on bridging different sensor domains with public communication networks and the Internet.

The existing literature describes a growing trend in NFV-based middlebox design. Since a WSN gateway falls under the taxonomy of *middlebox*, a brief overview of NFV architectures within the context of middleboxes is important. Therefore, we classify the state-of-the-art into two categories: Traditional Architectures (WSN/IoT gateway) and NFV architectures (middleboxes).

*1) Traditional Architectures (WSN/IoT Gateway):* An architecture for an in-home IoT gateway is proposed in [4]. It consists of three subsystems: Sensor node, gateway, and application platform. The architecture does not support standard or proprietary interfaces. Jiang et al. [5] present an IoT gateway architecture for a CorbaNet-based digital broadcast system, designed to lessen the effects of IoT technology on backbone networks. The architecture is extensible by nature. However, it doesn't account for information model conversion and its scalability aspect is not discussed.

A configurable, multifunctional and cost-effective architecture for smart IoT gateways is proposed in [6]. It is extensible since modules with different communication protocols can be plugged into the architecture. It also provides protocol conversion by granting a common frame structure for data communication. However, scalability in terms of number of applications is not discussed.

In [7], the authors propose an IoT gateway-centric architecture that provides various M2M services, such as association of metadata to sensor and actuator measurements using SenML. They also extend SenML capabilities to address actuator control. Although it is scalable in terms of handling traffic by using the RESTful paradigm, it cannot support the dynamic creation of additional M2M services with more IoT devices. In [8], gateway architecture for home and building automation system is proposed. The gateway is managed remotely by the network operator. The architecture supports standard and proprietary interfaces. However, scalability is not discussed.

*2) NFV architectures (Middleboxes):* ClickOS [9] is a Xen-based software platform that allows hundreds of middleboxes to run on commodity hardware. It includes both simple middleboxes (e.g., packet forwarding from input to output interfaces) and full-fledged middleboxes (e.g., IPv4 router, firewall, etc.). However, virtualizing gateway modules is not investigated. The architecture is scalable, flexible, and extensible.

T-NOVA [10] is an integrated architecture that enables network operators and service providers to manage their NFVs. It provides VNFs, like flow handling control mechanisms, as value-added services to its customers. T-NOVA allows third party developers to publish their VNFs as independent entities. In [11], NFV is used to virtualize an IP telephony function called a Session Border Controller (SBC), which operates on both the control plane (i.e., load balancing and call control) and the media plane (i.e., media adaptation capabilities). The use of NFV for the virtualization of routing function in OpenFlow-enabled networks is explored in [12]. These works neither target WSN/WSAN domains, nor support proprietary southbound interfaces.

We conclude that, with the exception of limited support for extensibility, proprietary interfaces, and gateway modules, the existing WSN gateway architectures fall short of satisfying many of our requirements. With regard to NFV-based solutions, the current NFV architectures for middleboxes exhibit extensibility and scalability properties. However, they focus primarily on network elements, e.g., firewall, proxies, and NATs.

III. PROPOSED NFV ARCHITECTURE FOR VIRTUALIZED WSAN GATEWAY

In this section, we present our NFV architecture for virtualizing WSAN gateways. The architectural principles are discussed first, followed by the architectural components and interfaces, VNF migration and scalability issues, control plane, and an illustrative scenario.

*A. Architectural Principles*

Our first architectural principle is granular provisioning of network functions. We aim to use highly granular VNFs for virtualized WSAN gateway functions. Examples include protocol conversion and information model conversion. The protocol converter decodes a packet received in one protocol and encodes it in another protocol. The information model conversion converts data from one format to another. We do acknowledge the fact that converting a protocol X (or an information model X) into a protocol Y (or an information model Y) is not always feasible. Consequently, the Gateway Provider provisions the related VNFs only when the conversion is feasible. Our second principle is that the VWSAN Gateway Provider maintains a centralized store of VNF images. VNFs are dispatched on-demand to the VWSAN provider's domain. This principle is in accordance with the ETSI, that VNFs must be deployed throughout the networks where they are most effective and highly customized to a specific application or user [13]. The third and last principle is that the interaction interfaces between different domains are REpresentational State Transfer (REST)-based. REST is selected because it is lightweight, standard-based, and can support multiple data representations (e.g., plain text, JSON, and XML).

*B. Overall Architecture*

Fig. 1 shows the proposed architecture. It comprises several Application Domains, a VWSAN Gateway Provider

Table 1-Resources on the VWSAN Provider Domain and VWSAN Gateway Provider Domain

| Domain Name | Resource | Operation | Http Action |
|---|---|---|---|
| Resources on VWSAN Provider Domain | List of application service requests | Create: Add application information (protocol used, data format, latency, etc.) | POST: /ApplicationsServiceRequests |
| | Specific application's service request | Update: Change information of specific application | PUT: /ApplicationsServiceRequests/(RequestId} |
| | | Delete: Delete specific application information | DELETE: /ApplicationsServiceRequests/(RequestId} |
| | Notification of service availability | Create: Send notification to VWSAN domain via the gateway domain about the availability of requested VNFs. | POST: /ServiceAvailabilityNotification |
| Resources on VWSAN Gateway Provider Domain | Request for VNFs | Create: Send request from VWSAN domain to gateway domain for VNFs with specific information (northbound interface, VWSAN description, etc.) | POST: /VNFsRequest |
| | Specific request for VNFs | Update: Change information of specific request for VNFs. | PUT: /VNFsRequest/{VNFsRequestId} |
| | | Delete: Delete information of specific request for VNFs. | DELETE: /VNFsRequest/{VNFsRequestId} |

Domain, and VWSAN Provider Domains. The components and interfaces are presented, followed by a discussion of the VNF migration and scalability issues.

*1) Components and Interfaces:*

*a) Components:* Each Application Domain contains an Application that requires the services of one or more VWSAN providers. The Application contains two components: Infrastructure Agent and Sensor/Actuator Agent. The Infrastructure Agent is responsible for the singaling procedure. It communicates with the VWSAN Provider Domain to negotiate the use of VWSAN infrastructure. The Sensor/Actuator Agent is responsible for gathering measurements from the sensor and sending commands to the robots. The VWSAN Gateway Provider Domain consists of the following entities:

- Core Layer: Contains VNFs and their corresponding Element Management Systems (EMS), where each EMS is responsible for monitoring the resource utilization of its corresponding VNF [13].

- NFV Infrastructure (NFVI): provides hardware and software resources, including computation, storage, and networking needed to deploy, manage, and execute VNFs.

- NFV Management and Orchestration (MANO): Responsible of orchestration and lifecycle management of physical/software resources, and the lifecycle management of VNFs (instantiation, update, migration, and termination).

- Central Controller: Performs functions as part of the signaling procedure that occurs during service negotiation (this is described later).

- VNF Store: A repository that contains VNFs of various gateway modules. It provides VNFs that match the requirements of an end-to-end service.

Each VWSAN Provider Domain comprises the following components:

- Southbound (SB) Handler Layer: Contains VNFs that have been migrated from the VWSAN Gateway Provider Domain and their corresponding EMSs.

- NFVI: (explained in previous section).

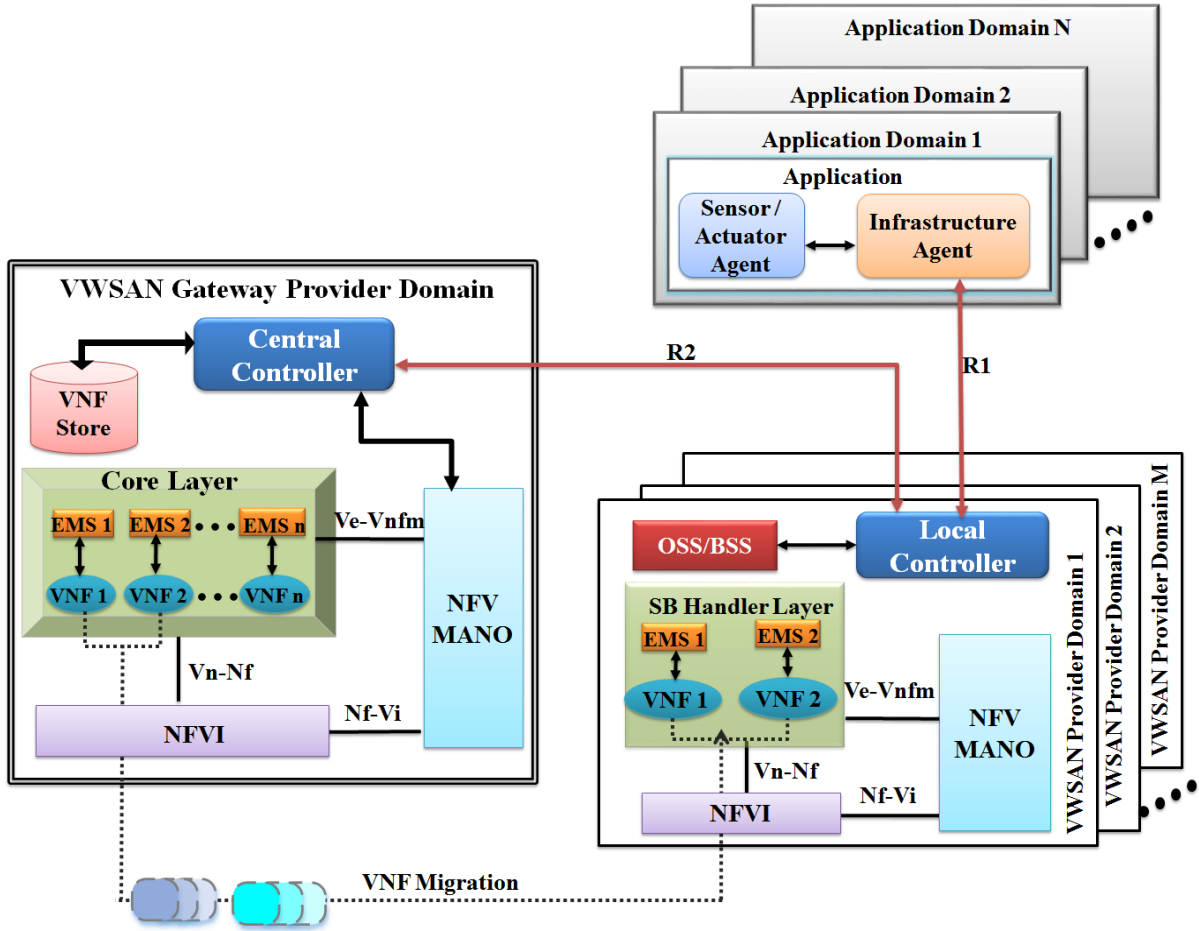

Figure 1. Overall Architecture

- NFV MANO: Performs the typical orchestration and management functions for the execution of migrated VNFs.
- Operational Support System/Business Support System (OSS/BSS): Provides the description of VWSAN (e.g., sensor/robot brands).
- Local Controller: Interacts with the Infrastructure Agent and the Central Controller.

  *b) Interfaces:* The NFV components i.e., Core Layer, NFVI, NFV MANO, SB Handler Layer interact with each other through the interfaces defined by ETSI [13]. They

  *a) VNF Migration:* In the architecture, VNFs are migrated on-demand from VWSAN Gateway Provider Domain to VWSAN Provider Domain. The architecture supports two approaches for migration. In the first approach, VNFs are instantiated and chained in VWSAN Gateway Provider Domain. Then, using live migration, running VMs are sent from the VWSAN Gateway Provider Domain to VWSAN Provider Domain. In the second approach, VNFs are migrated from the VWSAN Gateway Provider Domain to VWSAN Provider Domain, where they are instantiated and chained.

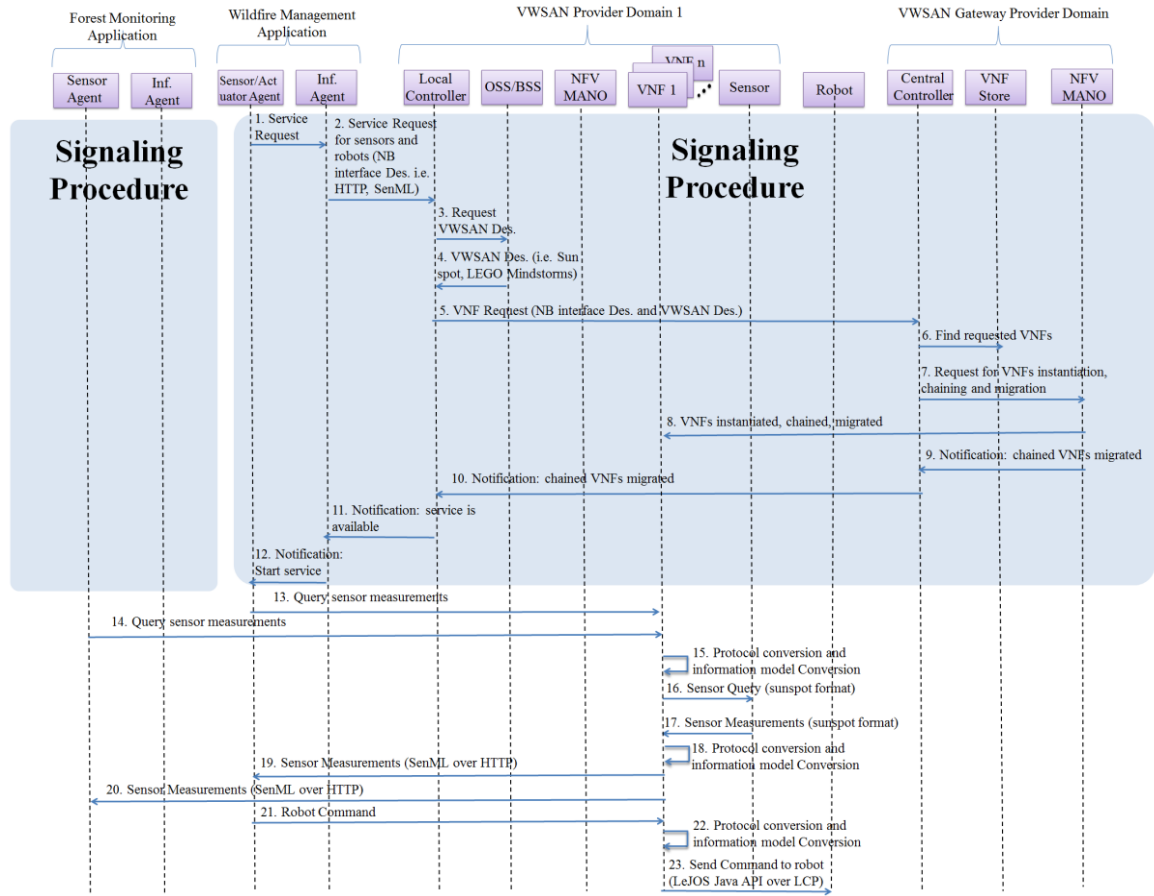

Figure 2. Sequence Diagram

include Vn-Nf, Nf-Vi and Ve-Vnfm. Vn-Nf represents the execution environment provided by NFVI to the Core Layer and to the SB Handler Layer. Nf-Vi is used for assigning virtualized resources in response to resource allocation requests (e.g., allocating VMs on hypervisors). It is also used by NFVI to communicate status information about virtualized and hardware resources to the MANO. Nf-Vi is also used to configure hardware resources. Ve-Vnfm carries out all operations during a VNF life cycle, including instantiation, scaling, updating, and termination. It is also used for exchanging VNF configuration information.

*2) VNF Migration and Scalability Issuses:*

*b) Scalability:* The architecture relies on dynamic resource allocation algorithms to meet the growing demand of applications. These algorithms enable vertical scaling – i.e., increasing the resources of a VNF instance (e.g., CPU, memory) and/or horizontal scaling – i.e., increasing the number of VNF instances that serve an application. Existing algorithms such as [14] and [15] could be used as basis. We consider the design of these algorithms as items for future work.

*C. Control Plane*

The control plane consists of signaling procedure and control interfaces, R1 and R2. In a typical end-to-end service, the application sends query to sensors to receive measurements and deploy robots. Before the service begins, a signaling procedure is conducted, in which different business players (i.e., Application Domain, VWSAN Provider, and VWSAN Gateway Provider) engage in service negotiation and exchange the necessary parameters to obtain the appropriate VNFs.

*1) Signaling procedure:* Signaling is initiated when an application requires services from VWSAN Provider Domain. The Sensor/Actuator Agent instructs the Infrastructure Agent to start the service negotiation. The Infrastructure Agent creates a service request that includes a description of the northbound interface used by the application (i.e., communication protocol, information model, etc.) and QoS parameters associated with the service delivery (i.e., latency, throughput, etc.) and sends it to the Local Controller of VWSAN Provider Domain. Upon receipt of the service request, the Local Controller communicates with the OSS/BSS to obtain informaton on parameters specific to the VWSAN (e.g., type of sensors/robots). It then creates a VNF request containing parameters of the service request as well as parameters specific to the VWSAN and sends it to the Central Controller. Based on these parameters, the Central Controller searches for appropriate VNFs in VNF Store.

If the VNFs are found, the Central Controller instructs NFV MANO of VWSAN Gateway Provider Domain to instantiate and migrate the VNFs to VWSAN Provider Domain. The Central Controller also receives a notification from NFV MANO of VWSAN Gateway Provider Domain when the VNFs are ready for use in VWSAN Provider Domain. The Central Controller then forwards the notification to the Local Controller, which sends a notification about service availability to the Infrastructure Agent, which then notifies the Sensor/Actuator Agent to start the service. It is important to note that, when the required VNFs are not found in the VNF Store, a service unavailability notification is sent to the Infrastructure Agent, to either cancel the negotiation or resume signaling after a certain time period.

*2) Control Interfaces:* R1 is used for the interactions between Infrastructure Agent and Local Controller. R2 is used for the interactions between Local Controller and Central Controller. R1 and R2 are based on REST paradigm. The important information is modelled as resources and each resource is uniquely identified by the Uniform Resource Identifier (URI). Table 1 summarizes the proposed REST interface for the interactions between different domains. It defines resources on VWSAN Provider Domain, used to reserve resources when it receives service request from Application Domain with a description of parameters. They also allow the Application Domain to modify parameters and delete resources of specific applications. Furthermore, they allow VWSAN Gateway Provider Domain to send notifications to VWSAN Provider Domain about the availability of the requested VNFs. The resources defined on VWSAN Gateway Provider Domain allow it to receive VNF requests from VWSAN Provider Domain. They also allow the VWSAN Provider Domain to update or delete information (e.g., sensor/robot brand) about specific VNF requests.

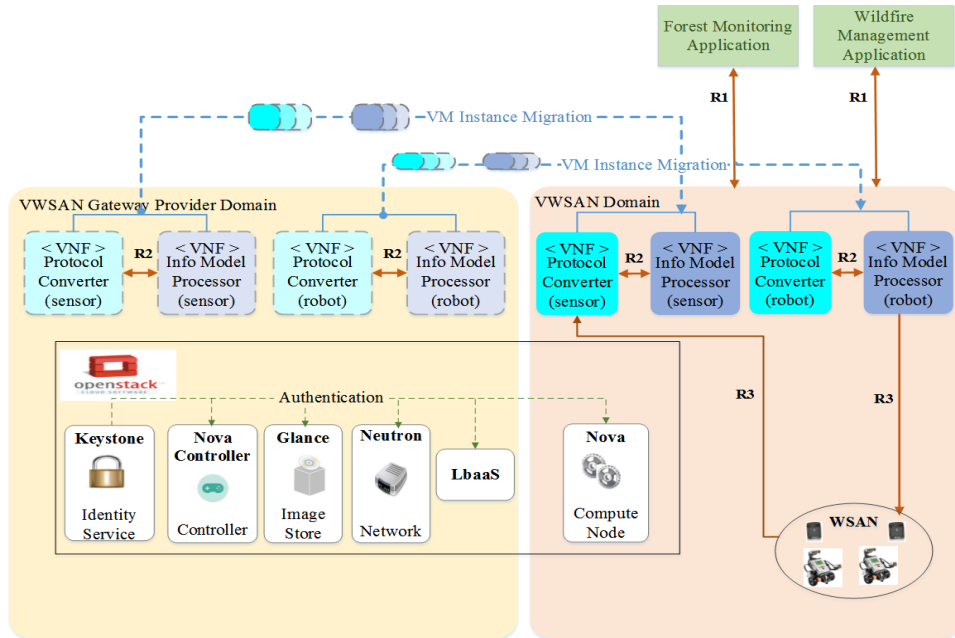

Figure 3. Prototype Architecture

*D. Illustrative Scenario*

In Fig. 2, we illustrate an end-to-end scenario, wherein an application (e.g., forest monitoring) queries the sensors owned by VWSAN Provider 1 and collect their measurements, and another application (e.g., wildfire management) needs to be notified when fire occurs and deploy robots. Before using VWSAN Provider Domain's service, the signaling procedure starts. The northbound interface description sent to the Local Controller for both sensors and robots is SenML over HTTP. Since the current SenML implementation only supports sensor measurements, we have used the extended capabilities of

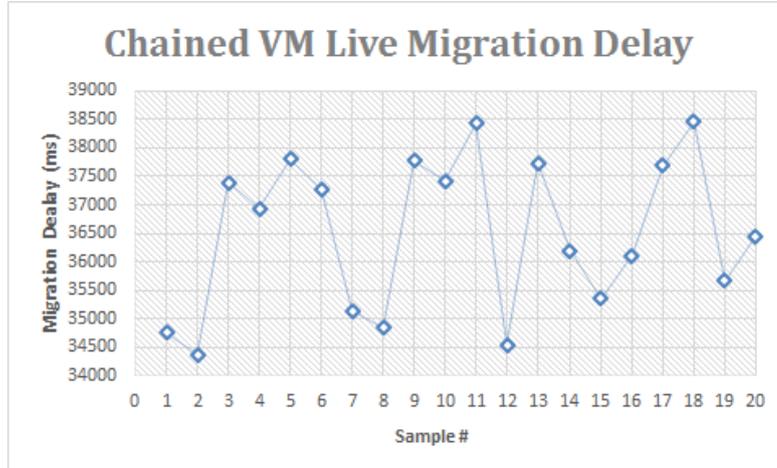

a)

b)

| Sample | Protocol Conversion Downtime (sec) | Info Model Processor Downtime (sec) |
|---|---|---|
| 1 | 30 | 39 |
| 2 | 24 | 39 |
| 3 | 31 | 38 |
| 4 | 33 | 38 |
| 5 | 24 | 38 |

c)

Figure 4. Results of Service Provisioning: a) Live migration delay of VM b) Pinging the VM during live migration c) VM downtime during live migration

SenML proposed by Datta et al. in [7] and [16] to send robot commands from the application.

Upon receiving the description from Infrastructure Agent, the Local Controller obtains a description of the sensors (i.e., SunSpot) and the robots (i.e., Lego Mindstorms) from OSS/BSS. The signaling procedure continues as described in section-III.C.1 for both applications. For VNF migration, the second approach (see section-III.B.3) is used; the VNFs are instantiated, chained, and then migrated to VWSAN Provider Domain. After service negotiation, the Sensor/Actuator Agent sends a query to the sensors through the VNFs. Upon receiving the query, SunSpot sensors send their raw measurements over CoAP protocol. These measurements are processed by protocol conversion (encoded in HTTP protocol) followed by information model conversion (mapped to SenML format), in order to enable the applications to interpret the measurements. If the wildfire management application receives notification of fire, it sends actuating commands to the robots in SenML format through HTTP, where the commands are mapped to LeJOS Java API and to Lego Communication Protocol (LCP). The end-to-end service is completed when the robots are deployed.

## IV. IMPLEMENTATION

### A. Prototype

For the prototype, we implemented the scenario in which the forest monitoring agency is interested in collecting environmental data to monitor the forests and a wildfire management agency that needs to be notified when fire occurs and deploy robots in order to suppress it. We consider a forest wherein WSANs have already been deployed to monitor and suppress wildfires. Two different brands sensors were used, each belonging to different WSAN cloud infrastructures. The sensors measure the temperature and can thereby detect fires and the robots can detect extinguisher and grab it in order to suppress the fire. In order to communicate with different types of sensors and robots, the application needs a gateway for handling different types of communication interfaces. A third party provider provides this gateway.

The forest monitoring and wildfire management applications was created using java dynamic web application and hosted on Tomcat8 server. We used OpenStack Icehouse to build our private cloud. OpenStack is a free, open-source software for creating private and public clouds. Fig. 3 depicts our prototype architecture. We used a multi-node OpenStack with two compute nodes. We considered each compute node as a domain: One as VWSAN Provider Domain and the other as VWSAN Gateway Provider Domain. In our prototype, we assume the two domains are in the same data center. In order to provide live migration, both compute nodes share the same storage. This allows the migration of only the memory footprint of the VM. If each domain were in a separate data center, we would assume a provision for live migration among them.

The VNFs are instantiated in VWSAN Gateway Provider Domain and migrated to VWSAN Provider Domain after being chained. For simplicity's sake, we assume that the VNFs are chained in a static way in VWSAN Gateway Provider Domain.

In the node representing VWSAN Gateway Provider Domain, all necessary components of Openstack were installed, including: Identity Service-Keystone, Controller-Nova, Image-Glance, and Networking-Neutron. NFS (Network File System) server was also configured in this node, allowing servers to share directories and files with each other over a network. The two nodes representing VWSAN Provider Domain contains Compute-Nova. The fourth node is configured as Network-Neutron and LBaaS (Load Balancing as a Service) was installed on it, which is a service of Neutron, allowing to load balance traffic for services running on VMs in OpenStack. We used OpenStack4j API, as an open-source OpenStack client, allowing the provision and control of an OpenStack system as a controller. Because all domains are in the same data centers, the controller can control all domains. Each VNF runs a Linux Ubuntu V14.04 on 1 VM, and is equipped with 1 VCPU and 2GB RAM. The VNFs communicate with each other through REST interface (R2), using the RESTlet framework [13]. Communication between VWSAN Provider Domain and Application Domains is also achieved via REST interface (R1).

### B. Setup

The applications and the domains controller run on a PC with Intel® Xeon® CPU clocked at 2.67 GHz and a 6GB RAM with 64-bit Windows 7 Enterprise. This PC uses JVM version 1.8.0_51. We used four PowerEdge™ T410s, which is an Intel® processor-based server – two as nova compute nodes, one as the nova controller, and one as the network node.

Two Java Sun SPOT sensors, two Advanticsys sensors, and one LEGO Mindstorms NXT robot were used. Each sensor executes the forest monitoring task. We implemented a simple gateway that runs on a laptop with Intel® Core ™i7-2620 CPU with 2.70Hz, and 8 GB of RAM. This gateway exposes the robots and the sensors capabilities as APIs. For example, in order to send command to the robot, the protocol converter and information model conversion convert the REST request received at its northbound interface to LeJOS Java API commands that implements the LCP. The gateway then wraps the request to either Bluetooth or USB communication channel and sends it to the robot.

### C. Performance Evaluations

*1) Performance Metrics:* The performance metrics according to which we evaluate system performance are:

*a) Service Provisioning Time:* Time between the moment the VM instantiation starts in VWSAN Gateway Provider Domain and when the VMs are migrated to VWSAN Provider Domain, including the chaining time of VMs, while also calculating the downtime duration of the VMs.

*b) End-to-End (E2E) Delay:* Time between the moment the sensors send a measurement and when the robots are

deployed. We calculated E2E delay for both, non-virtualized and virtualized environments.

*c) Scalability:* Ability of the system to handle the growing amount of loads without suffering significant degradation in the performance. We considered the *Response Time* of the system as a metric to evaluate the scalability of our architecture. Response time is the time period from when measurements are sent by the sensors to when these measurements are received by the VNFs.

*2) Results and Discussions:* This section discusses the performance results obtained, beginning with the live migration delay.

*Test Case 1: Service Provisioning Time*

Fig. 4-a depicts the live migration delay of chained VMs, based on shared storage in a virtualized environment. We studied 20 tests and found a maximum delay of 38.4s and a minimum delay of 34.3s. We observed that the delay fluctuates between samples. This is because the time needed to instantiate VMs and migrate them in OpenStack is inconsistent. One of the limitations of OpenStack is the time needed to start a new VM, which could cause a prolonged delay in service provisioning time. As reported in [17], VM instantiation delay can sometimes reach up to 60s.

Although the live migration of VMs allows to transfer VM to other physical servers without shutdown and ensures high availability with non-stop services, VMs still face some period of downtown, depending on the memory state of the VM. In this experiment, we tested ping on the VMs during live migration.

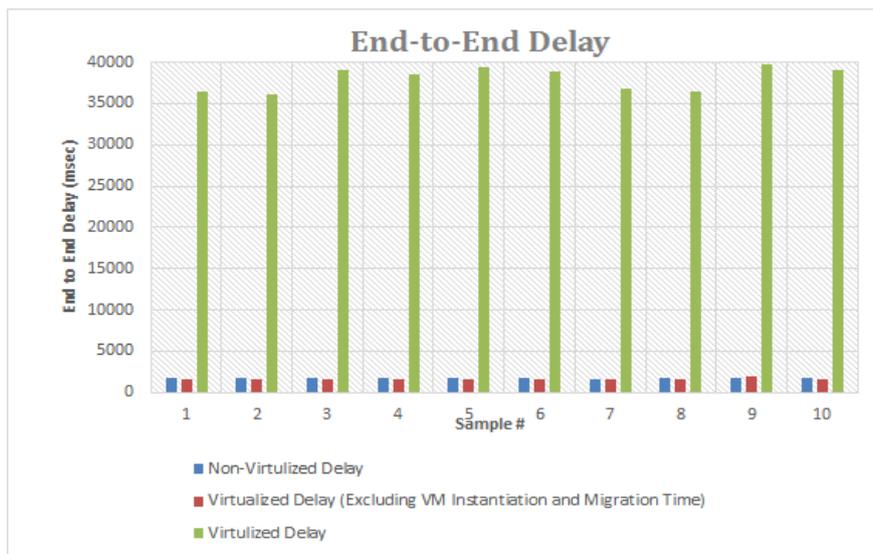

a)

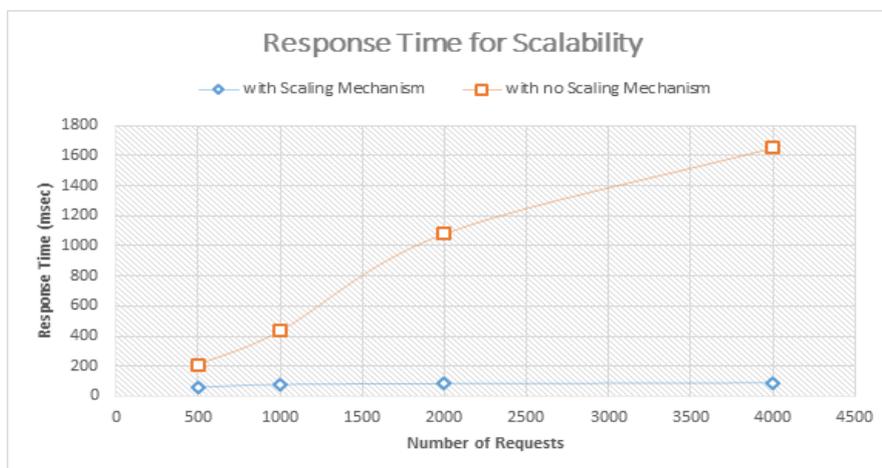

b)

Figure 5. Results: a) End to end delay (virtualized gateway vs. non-virtualized gateway) b) Response time for scalability

We started pinging before the live migration starts and it lasted until the migration ends. We noticed that during the migration, some ping requests were lost. Fig. 4-b shows the process of pinging the VMs and Fig. 4-c shows the downtime of the VMs considering 5 samples.

*Test Case 2: E2E Delay*

Fig. 5-a illustrates a comparison of E2E delay for virtualized and non-virtualized environments, wherein each sample represents the average E2E delay for the first 10 measurements. In order to ensure an accurate comparison, we repeated the experiment 10 times. The average E2E delay for virtualized environment is always higher than the delay for non-virtualized environment. This is because the delay includes the time needed to instantiate and migrate the VMs. The maximum E2E delay for the virtualized gateway is around 39736 msec (sample 9), whereas it is 1784 msec (sample 10) for the non-virtualized gateway. Frameworks such as [18] can be integrated with OpenStack to overcome the performance gap between virtualized and non-virtualized environments.

We observe that the minimum E2E delay of virtualized gateway, excluding VM instantiation and migration delay, (1603 msec) is close to the minimum E2E delay of non-virtualized gateway (1601ms). Thus, we can conclude that the time needed to instantiate and migrate the chained VMs has a significant impact on E2E delay of virtualized gateway, which demonstrates the overhead of virtualization. However, E2E delay in a virtualized environment increases only when a new brand of sensor joins and sends requests to dispatch VMs to VWSAN Gateway Provider Domain.

*Test Case 3: Scalability*

We used a simplified resource allocation mechanism to test the scalability. It is based on the resource utilization of the VMs (i.e., CPU) and on horizontal scaling. Each T period of time, VM's resources are monitored. If the utilization of the resource exceeds the threshold (i.e., 70%), we perform horizontal scaling. To conduct our case study, we set the number of requests as variable within a unit time (T) and gradually increase it from 500 to 4000 requests. We considered 10sec as the unit time. We used Apache JMeter to generate the requests using a uniform distribution of threads.

Fig. 5-b shows the results of our experiments, where we compared it with the same scenario without having a scaling mechanism. In the case of having scaling mechanism, we notice that as the load (i.e., number of requests) increases, the system experiences a very slight increase in response time. This is because scaling is triggered before the system enters the overload state. For the initial increase in load (i.e., from 500 to 1000), the effect on response time is slightly more than the one afterwards. This is because initially as load increases, more resources cannot be allocated until the T period is elapsed. From load 1000 till the maximum load, the response time increases by only 5ms for every 2-fold increase in load. In contrast, if no scaling is performed, the system suffers from a significant increase in response time, as indicated in the figure. We observe that from load 1000 till the maximum load, the response time increases by 600ms for every 2-fold increase in load. Overall, with a scaling mechanism, the load has a very negligible impact on the response time. This demonstrates the scalability of our architecture.

## V. CONCLUSION AND FUTURE WORKS

In this paper, we introduce an NFV architecture that deploys virtualized instances of a VWSAN gateway in an NFV infrastructure. The virtualized instances are dynamically migrated from a Gateway Provider Domain to several VWSAN Domains. With NFV, it is possible to achieve scalable deployment of gateways in heterogeneous VWSAN environments. In addition, several business actors involved in the proposed NFV architecture creates potentials for unique business models.

We also discuss a proof-of-concept of the NFV-based virtualized gateway. We evaluate the prototype by conducting a set of experiments. The performance comparison of virtualized and non-virtualized approaches is analyzed, and the scalability of the architecture is proved.

There are several potential items for future work. An example is the host of security and trustability issues brought by the introduction of the VWSAN gateway provider (or more generally new actors). Another example is the distribution of virtualized environment in the VWSAN domain. New interface mechanisms will then be required between the gateway provider and the different nodes that will host the VNFs in the distributed virtualized environment and also between the VNFs that now reside on separate nodes in this very same environment. Standardization will indeed be required to ensure interoperability. Yet another example is the design of resource allocation algorithms in the specific context of VNFs. A potential starting point is the resource allocation algorithms that exist today for VMs.

### ACKNOWLEDGMENT

This work is partially supported by CISCO systems through grant CG-576719.

BIOGRAPHIES

**Carla Mouradian** received her Bachelor's degree in Telecommunication Engineering from University of Aleppo, Syria in 2009, and obtained her Master's degree in Electrical and Computer Engineering from Concordia University, Canada in 2014. She is working towards her Ph.D. degree in Information System Engineering at Concordia University. Her research interests include cloud computing, wireless sensor networks, Network Function Virtualization, and Internet of Things. She is a member of the IEEE Communications Society.

**Tonmoy Saha** is currently pursuing his Master of Computer Science from Concordia University, Montreal, Quebec, Canada and received his B.Sc (Hons) in Computer Science & Engineering from Jahangirnagar University, Savar, Dhaka, Bangladesh. He worked as a Senior Software Engineer in Solution Lab at Samsung R&D Institute Bangladesh. His research interests are Cloud Computing, Wireless Sensor Network, Internet of Things, Network Function Virtualization and Software Engineering.

**Jagruti Sahoo** received a Ph.D. degree in computer science and information engineering from the National Central University, Taiwan, in January 2013. She worked as Postdoctoral Fellow in University of Sherbrooke, Canada from 2013 to 2014. She is currently a Postdoctoral Fellow at the Telecommunication Service Engineering Research Laboratory, CIISE, Concordia University, Canada. Her research interests include wireless sensor networks, vehicular networks, content delivery networks, Cloud Computing and Network Functions Virtualization. She served as a member of the Technical Program Committee in many conferences and as a Reviewer for many journals and conferences. She is a member of the IEEE Communications Society.

**Mohammad Abu-Lebdeh** received his B.Sc. degree in Computer Engineering from An-Najah National University, Palestine, and M.Sc. degree in Electrical & Computer Engineering from Concordia University, Canada. He is currently pursuing his Ph.D. degree in Information & Systems Engineering at Concordia University. In the past, he worked for several years as a software engineer. His current research interests include cloud computing, service engineering, and next generation networks.

**Roch Glitho** holds a Ph.D. (Tekn. Dr.) in tele-informatics (Royal Institute of Technology, Stockholm, Sweden), and M.Sc. degrees in business economics (University of Grenoble, France), pure mathematics (University of Geneva, Switzerland), and computer science (University of Geneva).

He is an associate professor and Canada Research Chair at Concordia University. He is also an adjunct professor at several other universities including Telecom Sud Paris, France, and the University of Western Cape, South Africa. In the past, he has worked in industry and has held several senior technical positions (e.g., senior specialist, principal engineer, expert) at Ericsson in Sweden and Canada. His industrial experience includes research, international standards setting, product management, project management, systems engineering, and software/firmware design. He has also served as an IEEE Distinguished Lecturer, Editor-In-Chief of IEEE Communications Magazine, and Editor-In-Chief of IEEE Communications Surveys & Tutorials Journal.

**Monique Morrow** holds the title of CTO Cisco Services. Ms. Morrow's focus is in developing strategic technology and business architectures for Cisco customers and partners. With over 13 years at Cisco, Monique has made significant contributions in a wide range of roles, from Customer Advocacy to Corporate Consulting Engineering. With particular emphasis on the Service Provider segment, her experience includes roles in the field (Asia-Pacific) where she undertook the goal of building a strong technology team, as well as identifying and grooming a successor to assure a smooth transition and continued excellence. Monique has consistently shown her talent for forward thinking and risk taking in exploring market opportunities for Cisco. She was an early visionary in the realm of MPLS as a technology service enabler, and she was one of the leaders in developing new business opportunities for Cisco in the Service Provider segment, SP NGN. Monique holds 3 patents, and has an additional nine patent submissions filed with US Patent Office. Ms. Morrow is the co-author of several books, and has authored numerous articles. She also maintains several technology blogs, and is a major contributor to Cisco's Technology Radar, having achieved Gold Medalist Hall of Fame status for her contributions. Monique is also very active in industry associations. She is a new member of the Strategic Advisory Board for the School of Computer Science at North Carolina State University. Monique is particularly passionate about Girls in ICT and has been active at the ITU on this topic - presenting at the EU Parliament in April of 2013 as an advocate for Cisco. Within the Office of the CTO, first as an individual contributor, and now as CTO, she has built a strong leadership team, and she continues to drive Cisco's globalization and country strategies.

**Paul Polakos** is currently a Cisco Fellow and member of the Mobility CTO team at Cisco Systems focusing on emerging technologies for future Mobility systems. Prior to joining Cisco, Paul was Senior Director of Wireless Networking Research at Bell Labs, Alcatel-Lucent in Murray Hill, NJ and Paris, France. During his 28 years at Bell Labs he worked on a broad variety of topics in Physics and in Wireless Networking Research including the flat-IP cellular network architecture, the Base Station Router, femtocells, intelligent antennas and MIMO, radio and modem algorithms and ASICSs, autonomic networks and dynamic network optimization. Prior to joining Bell Labs, he was a member of the research staff at the Max-Planck Institute for Physics and Astrophysics (Munich) and visiting scientist at CERN and Fermilab. He holds BS, MS, and Ph.D. degrees in Physics from Rensselaer Polytechnic Institute and the University of Arizona, and author of more than 50 publications and 30 patents.